\documentclass[aps,prb,twocolumn,groupedaddress,showkeys,showpacs]{revtex4}
\usepackage{multirow,amssymb,amsbsy,amsmath}
\usepackage{epsfig}

\usepackage{graphicx}% Include figure files
\usepackage{dcolumn}% Align table columns on decimal point
\usepackage{bm}% bold math

\newcommand{\figref}[1]{Fig.~\protect\ref{#1}}
\begin{document}

\title{Sampling the two-dimensional density of states $g(E,M)$ of a giant magnetic molecule using the Wang-Landau method}% Force line breaks with \\

\author{Stefan Torbr\"ugge}
\email{storbrue@uos.de}
\author{J\"urgen Schnack}
\affiliation{%
Fachbereich Physik, Universit\"at Osnabr\"uck, Barbarastra{\ss}e
7, D-49069 Osnabr\"uck, Germany
}%

\date{\today}% It is always \today, today,

\begin{abstract}
  The Wang-Landau method is used to study the magnetic
  properties of the giant paramagnetic molecule
  \{Mo$_{72}$Fe$_{30}$\} in which 30 Fe$^{3+}$ ions are coupled
  via antiferromagnetic exchange. The two-dimensional density of
  states $g(E,M)$ in energy and magnetization space is
  calculated using a self-adaptive version of the Wang-Landau
  method. From $g(E,M)$ the magnetization and magnetic
  susceptibility can be calculated for any temperature and
  external field.
\end{abstract}


\maketitle

%=================================================================
\section{\label{Introduction}INTRODUCTION}

During the past decade the field of molecular magnets has
experienced a rapid evolution.\cite{GCR:S94,CGS:JMMM99,GSV:2006}
Nowadays a vast variety of species can be synthesized, ranging in
size from 2 to more than 30 paramagnetic ions embedded in the host
molecule.\cite{MPP:CR98}  The fascinating properties of these new
materials include hysteretic
behavior,\cite{CGS:JMMM99,CWM:PRL00,WKS:PRL02}  quantum tunneling
of the magnetization,\cite{TLB:Nature96,FST:PRL96,GaS:ACIE03}
magnetocaloric,\cite{WKS:PRL02,Sch:JLTP06} and magnetostrictive
effects.\cite{SBL:PRB06}

Astonishingly, for several observables and not too low
temperatures it is sufficient to treat magnetic molecules
classically by applying the Heisenberg model.\cite{Muller2001}
This enables one to apply the powerful machinery of classical
stochastic sampling methods.  The Metropolis algorithm
\cite{MMR:JCP53} has become the standard tool to calculate
statistical classical properties of these nano
magnets.\cite{Schroder2005_2,Schroder2005} Recently, the
Wang-Landau algorithm\cite{Wang2001} (WL) has been successfully
applied to various problems in statistical physics and biophysics
both to models with discrete \cite{Wang2001,Wang2001_2} as well as
with continuous degrees of
freedom.\cite{Kim2006,Rathore2004,Yan2003,Zhou2006} In this
context the WL exhibits a superior feature in comparison to the
Metropolis algorithm.  Since the WL calculates the density of
states (DOS), one can estimate thermodynamic observables such as
the free energy and entropy for all temperatures using just one
single simulation of the density of states.

However, the calculation of the DOS in energy space $g(E)$ permits
only to calculate thermodynamic properties as a function of
temperature and at zero magnetic field, e.g. the zero-field
specific heat, but not observables such as the magnetic
susceptibility at non-vanishing field.\cite{Brown2005}  For
studying the magnetic properties of a system at arbitrary external
magnetic field one has to calculate a joined DOS $g(E,M)$ in
energy and magnetization space. Once $g(E,M)$ is known, properties
like magnetization $M$ and magnetic susceptibility $\chi$ can be
calculated at any temperature and any external magnetic field
using again just one single simulation of the density of
states.\cite{Zhou2006,Zhou2006_2}

In this article we introduce a self-adaptive version of the WL
which allows to calculate the two-dimensional DOS $g(E,M)$ using a
discrete binning scheme in the continuous energy and magnetization
space. We demonstrate that the WL is capable to calculate
efficiently all thermodynamic properties of rather large spin
systems using the example of the magnetic Keplerate molecule
\{Mo$_{72}$Fe$_{30}$\}.\cite{Muller1999} The dependencies of the
thermodynamic observables both on temperature and external
magnetic field can be obtained by only one single simulation. We
conclude by discussing the problems arising at low temperatures
and high magnetic fields.

%=================================================================
\section{\label{ModelMethod}MODEL AND COMPUTATIONAL METHODS}

\begin{figure}[ht!]
\centering
\includegraphics[clip,width=40mm]{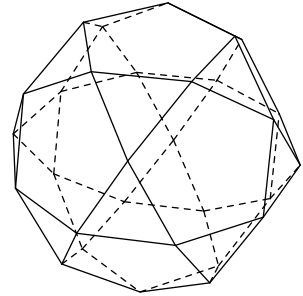}
\caption{Geometrical structure of an icosidodecahedron. For the
  molecule \{Mo$_{72}$Fe$_{30}$\} the vertices represent spin
  sites and the edges represent nearest neighbor interactions.}
\label{F-A}
\end{figure}

In the Keplerate molecule \{Mo$_{72}$Fe$_{30}$\} 30 iron~(III)
ions ($s=5/2$) occupy the vertices of a perfect
icosidodecahedron,\cite{Muller1999} see \figref{F-A}. The spins
are coupled with their nearest neighbors by an isotropic and
antiferromagnetic coupling of strength $J/k_B=1.566$~K. The
spectroscopic splitting factor is $g=1.974$.\cite{Muller2001}

%===================    figure   =================================
%===================    figure   =================================

We write the Heisenberg Hamiltonian as
\begin{equation}\label{Hamiltonian}
H = J \sum_{\langle m,n\rangle} \mathbf{S}_m \cdot \mathbf{S}_n +
g\mu_{B}{B}^{(z)}\cdot\sum_n {S}^{(z)}_n \ ,
\end{equation}
whereas $\langle m,n\rangle$ directs that the sum is over distinct
nearest-neighbor pairs, ${B}^{(z)}$ is an external magnetic field
in $z$-direction, $\mu_B$ is the Bohr magneton, and $\mathbf{S}$
denotes classical vectors of length
$|\mathbf{S}|=\sqrt{(s(s+1))}$.\cite{Luscombe1998}  We use a
self-adaptive \cite{Troster2005} scheme of the WL to calculate the
two-dimensional DOS $g(E,M)$, where $E$ denotes the Heisenberg
energy of a given spin configuration of Hamiltonian
(\ref{Hamiltonian}) without external magnetic field, i.e.
\begin{equation}\label{EHamiltonian}
E = J \sum_{<m,n>} \mathbf{S}_m \cdot \mathbf{S}_n \ ,
\end{equation}
and $M$ is the magnetization in $z$-direction, which is defined as
the sum over the $z$-components of all classical spin vectors:
\begin{equation}\label{MEq}
M = \sum_n {S}^{(z)}_n \ .
\end{equation}
In contrast to the Metropolis algorithm, where the acceptance
probability of a new generated state is determined by
$\min\{\exp[-(E_j-E_i)/(k_{\mathrm{B}}T)],1\}$, the WL is
characterized by an acceptance ratio
$\min\{g(E_{j})/g(E_{i}),1\}$, where $E_i$ and $E_j$ refer to
energies before and after the transition. The original WL has been
applied to models where the energy assumes only discrete
values.\cite{Wang2001,Wang2001_2} Since the Heisenberg model
consists of classical spin vectors which are continuously
orientable, the possible energy and magnetization values are real
numbers between the minimum and maximum energy and magnetization,
respectively. Thus, we discretize the continuous energy and
magnetization range by the introduction of
bins.\cite{Brown2005,Zhou2006} We have chosen bins of uniform
width of $(\Delta E)/k_{\mathrm{B}} = 4$~K in energy and $\Delta M
= 0.6\cdot|\mathrm{S}|$ in magnetization range. To speed up the
simulation we divide the total energy range in eight overlapping
intervals and follow the recipe in Ref.~\onlinecite{Schulz2003} to
avoid boundary effects. A priori it is not known, which
magnetization bins are accessible by the random walker in a
certain energy interval. A striking feature of the WL is, that one
does not have to know anything about the DOS one wants to
calculate. Following the original WL, the random walker biases
itself to explore the accessible energy and magnetization space.
To do so, we have chosen the following procedure: At the beginning
we perform a trial run for the desired energy interval. We
introduce an initial DOS $g_{initial}(E,M)=1$ and a reference
histogram $RH(E,M)$, and perform spin flips according the original
WL procedure accepting and neglecting new spin configurations for
a defined number of Monte-Carlo steps $n^{MC}_{initial}$. One
Monte-Carlo step corresponds to a single spin tilt event. Each
time a bin ($E_i,M_i$) is visited the corresponding entries in
$RH(E_i,M_i)=RH(E_i,M_i)+1$ and $g_{initial}(E_i,M_i)=f\cdot
g_{initial}(E_i,M_i)$ are updated, whereas $f$ is the initial
modification factor. Following this recipe the random walker
triggers itself to explore the accessible energy and magnetization
space for a predefined energy interval.

%===================    figure   =================================
\begin{figure}[ht!]
\centering
\includegraphics[clip,width=60mm]{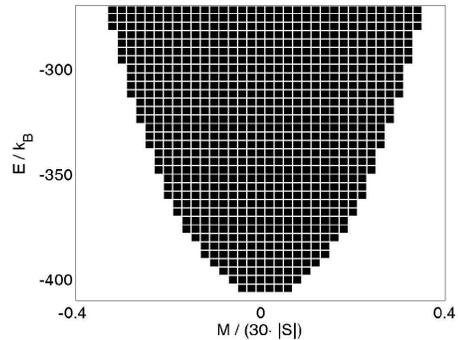}
\caption{Low-energy part of the two dimensional map for the
  reference histogram with $RH(E,M)\ne 0$ for the molecule
  \{Mo$_{72}$Fe$_{30}$\}. $RH(E,M)\ne 0$ defines the accessible
  energy and magnetization space. The grid shows the bins in energy and magnetization space.}
\label{F-B}
\end{figure}
%===================    figure   =================================

After $n^{MC}_{initial}$ steps have been performed we stop the
initial run. Now we continue with the normal WL for the same
energy interval. As an initial guess for the DOS we use
$g_{initial}(E,M)$. A new state is only accepted if the
corresponding entry in $RH(E,M)$ is not zero, compare
\figref{F-B}. In other words, we accept only states corresponding
to bins in the energy and magnetization space which have already
been visited during the initial run, otherwise the new generated
state is neglected according to Ref.~\onlinecite{Schulz2003}. In
addition, the relative flatness of the accumulated histogram is
only checked after all valid entries in the reference histogram
have been visited at least one hundred times. Of course the total
runtime of the simulation as well as the accuracy of the finally
calculated DOS are very sensitive to a good choice of
$n^{MC}_{initial}$. If $n^{MC}_{initial}$ is chosen very large,
the random walker explores bins at the boundaries of the
accessible energy and magnetization space, which are rarely
visited. Thus the DOS for these entries is small and does not
contribute significantly to the partition function.  Nevertheless,
sampling of these states increases the total runtime of the
simulation since a flat histogram is to be accumulated during the
simulation. On the other hand, if $n^{MC}_{initial}$ is chosen too
small, the energy and magnetization space is sparsely explored and
many bins of the accessible energy and magnetization space are not
visited, resulting in an inaccurate estimate of the DOS. Thus, the
question is to find a good tradeoff between runtime and accuracy
of $g(E,M)$. To estimate a reasonable value for $n^{MC}_{initial}$
for each energy interval, we have chosen different values and
counted the number of visited distinct bins in $RH(E,M)$ after
finishing the initial run. It turns out that the number of visited
bins shows some saturation behavior. After a critical number of
steps the visited energy and magnetization space does not grow
significantly any more. Thus, after this critical number of steps
has been reached we consider the accumulated $RH(E,M)$ to be a
good estimate for the successive WL run. As the initial
modification factor $f$ we have chosen a rather large value of
$f_{start}=e^4$ and reduce $f$ in large steps.\cite{Zhou2005}
After finishing the initial run we perform 14 steps decreasing $f$
according to the recipe $f_{i+1} = \sqrt[4]{f_i}$ resulting in a
final modification factor of 1.000000015. After $g(E,M)$ has been
obtained, the magnetization $M(T,B)$ as a function of temperature
$T$ and external magnetic field $B$ can be calculated from
\begin{equation}\label{EqMagn}
    \langle M(T,B)\rangle = \frac{\sum_{E,M} M g(E,M)
    \exp(-H/k_{\mathrm{B}}T)}{\sum_{E,M}g(E,M)\exp(-H/k_{\mathrm{B}}T)}.
\end{equation}
The differential magnetic susceptibility $\chi=\partial \langle
M\rangle/{\partial B}$ can be computed by using the equation
\begin{equation}\label{EqSusz}
    \chi(T,B) = \frac{1}{k_{\mathrm{B}}T}[\langle M^2(T,B)\rangle-\langle
    M(T,B)\rangle^2].
\end{equation}

While without external magnetic field it is sufficient to
calculate only the DOS for negative energies, since states with
positive energies due to their Boltzmann factor
$\exp(-E/k_{\mathrm{B}}T)$ practically do not contribute to the
partition function, in the case of an external magnetic field also
energy bins with a positive energy become important, since they
might be low-lying in the case of an applied external field. Thus
it is required to calculate the DOS over the full energy range of
the system, if the properties at high magnetic fields are studied.
The exact ground state energy for \{Mo$_{72}$Fe$_{30}$\} is known
to be $E_0/k_{\mathrm{B}} = -412.125$~K,\cite{Axenovich2001} which
is close to the lowest accessible energy bin of
$E_{min}/k_{\mathrm{B}}=-406$~K. The classical ground state of the
molecule is charactereized by relative angles of 120$^\circ$
between nearest-neighbor spins.\cite{Axenovich2001} The
possibility of generating such a spin configuration is practically
zero, and thus an effective sampling of the DOS near the ground
state energy is not possible. The same is valid for the highest
energy states of the molecule, where the probability of generating
a spin configuration with all spins pointing in the same direction
is unlikely. The highest energy bin visited during the initial run
is $E_{max}/k_{\mathrm{B}}=786$~K. As the criterion of flatness we
used a relatively small value of 0.5, and to further decrease the
statistical error we perform four runs, calculate the magnetic
properties for each of these runs and compute the mean value and
mean standard deviation afterwards. On a personal computer (Intel,
Xeon, 2.60~GHz) the sampling of the complete DOS took about 120
hours of cpu time.

For a comparison of accuracy we performed extensive Monte-Carlo
simulations using the Metropolis algorithm. We perform $30\cdot
10^6$ spin tilt trials to let the system reach equilibrium at a
defined external magnetic field and temperature. Afterwards for
additional $30\cdot 10^6$ tilting trials the thermodynamic
properties are computed.

%=================================================================
\section{\label{ResDis}RESULTS AND DISCUSSION}

%===================    figure   =================================
\begin{figure}[ht!]
\centering
\includegraphics[clip,width=0.8\columnwidth]{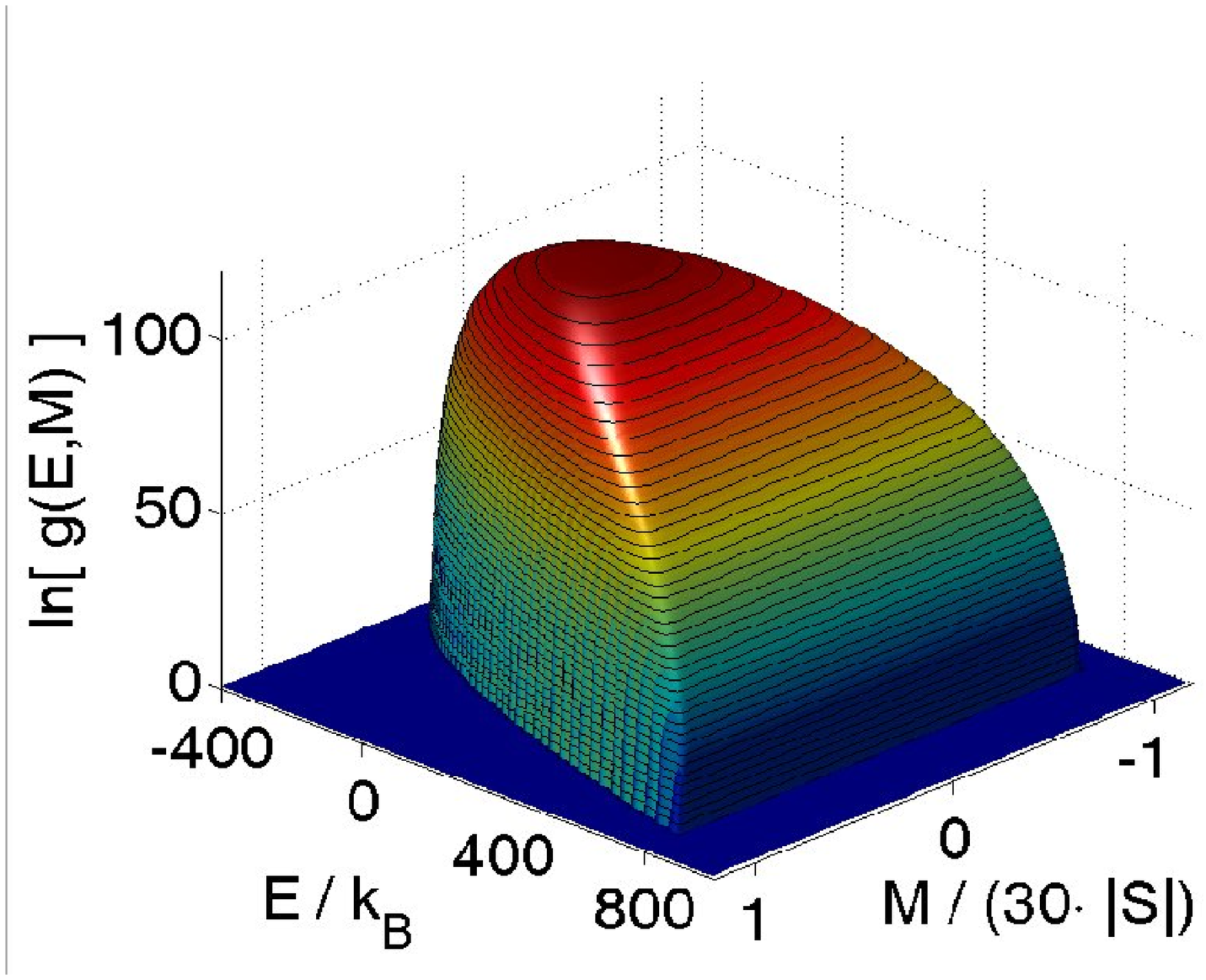}

\includegraphics[clip,width=0.8\columnwidth]{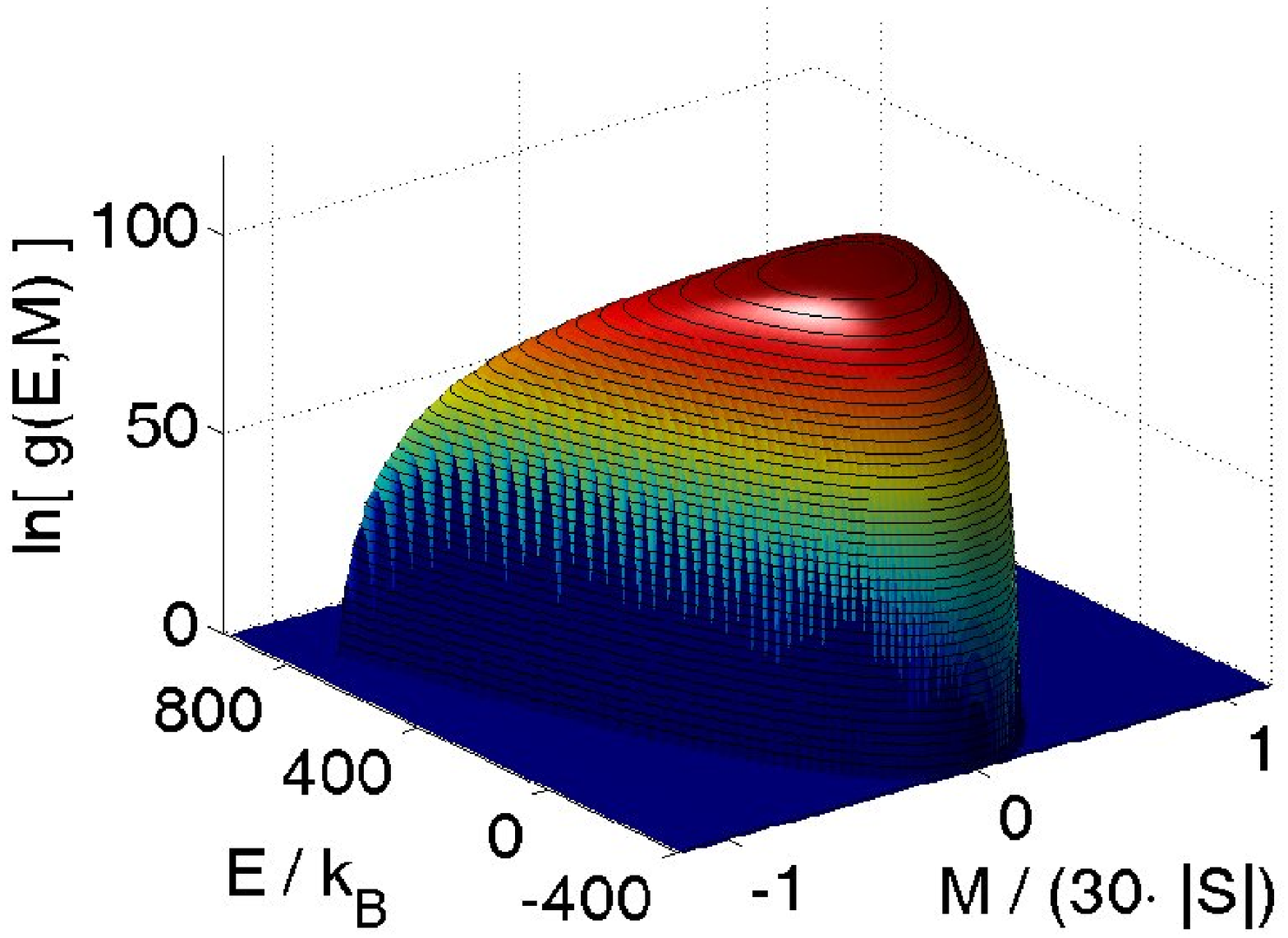}
\caption{\label{2D_DOS} (Color online) Joint DOS $\ln[g(E,M)]$ for
the magnetic molecule \{Mo$_{72}$Fe$_{30}$\}.}
\end{figure}
%===================    figure   =================================
The calculated joint DOS $g(E,M)$ for the magnetic molecule
\{Mo$_{72}$Fe$_{30}$\} is presented in Fig.~\ref{2D_DOS}. One
notices that the low-energy boundary $E_{\text{min}}(M)$ assumes
a parabolic shape as observed for various
systems.\cite{Wal:PRB01,ScL:PRB01} The region with $\partial
g(E,M)/\partial M = 0$ at high energies indicates the global
rotational symmetry of the system. One also notices that in
the vicinity of the phase-space boundary the classical density
of states grows very rapidly by orders of magnitude which
explains why it is difficult to obtain very accurate results at
the boundaries.

%===================    figure   =================================
\begin{figure}[ht!]
\centering
\includegraphics[clip,width=0.8\columnwidth]{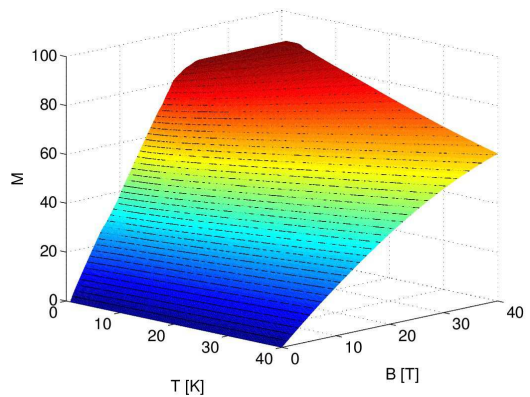}
\caption{\label{MvsTB} (Color online) Magnetization $M$ as a
function of temperature $T$ and external magnetic field $B$.}
\end{figure}
%===================    figure   =================================

The limited accuracy at the energy and magnetization space
boundaries looses its significance with increasing temperature.
This is demonstrated by evaluating the magnetization as a function
of temperature $T$ and external magnetic field $B$.
Figure~\ref{MvsTB} shows the behavior of the magnetization in the
relevant temperature and field range. The simulated magnetization
does not show any significant statistical fluctuations; it
compares nicely to the result of a Metropolis sampling.

%===================    figure   =================================
\begin{figure}[ht!]
\centering
\includegraphics[clip,width=0.6\columnwidth]{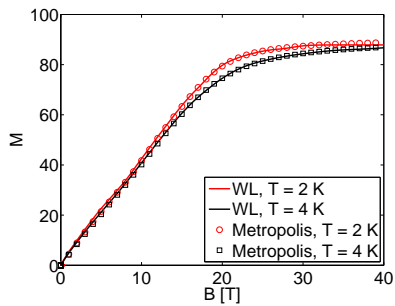}
\caption{\label{M_WL_Metr_Theory} (Color online) Comparison of the
magnetization $M$ as a function of external magnetic field
calculated from $g(E,M)$ (solid lines) with computed data using
the Metropolis algorithm at $T = 2$~K (open circles) and 4~K (open
squares), respectively. Statistical sampling errors are smaller
than the used symbols and line widths.}
\end{figure}
%===================    figure   =================================

Figure~\ref{M_WL_Metr_Theory} compares the magnetization at $T =
2$~K and 4~K as a function of field for the WL and the Metropolis
simulations. As can be inferred from the figure, at a temperature
of the order of the exchange coupling the WL reaches the same
accuracy as the Metropolis algorithm, but the latter has to be
performed for each pair of variables $(T,B)$, whereas with the WL
the density of states has to be sampled only once in order to
obtain the complete function $ \langle M(T,B)\rangle$.

%===================    figure   =================================
\begin{figure}[ht!]
\centering
\includegraphics[clip,width=0.8\columnwidth]{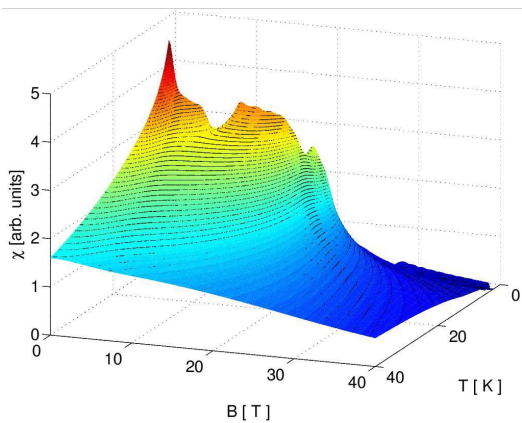}
\caption{\label{SuszvsTB} (Color online) Magnetic susceptibility
$\chi$ as a function of temperature $T$ and external magnetic
field $B$.}
\end{figure}
%===================    figure   =================================

%===================    figure   =================================
\begin{figure}[ht!]
\centering
\includegraphics[clip,width=0.6\columnwidth]{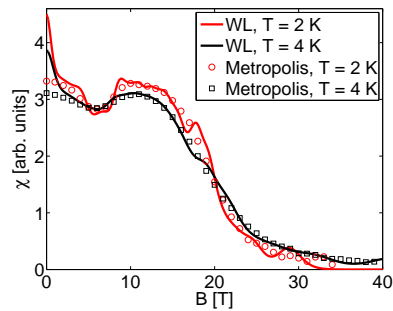}
\caption{\label{SuszvsB} (Color online) Comparison of the magnetic
susceptibility $\chi$ as a function of external magnetic field
calculated from $g(E,M)$ (solid lines) with computed data using
the Metropolis algorithm at $T = 2$~K (open circles) and 4~K (open
squares), respectively.}
\end{figure}
%===================    figure   =================================

The differential susceptibility, Figs.~\ref{SuszvsTB} and
\ref{SuszvsB}, is a second derivative of the partition function.
Therefore, it will magnify inaccuracies of the simulated density
of states. Figure~\ref{SuszvsTB} displays the susceptibility as
a function of $T$ and $B$. It can be seen that for temperatures
higher than $J/k_B$ the behavior is rather smooth, but for lower
temperatures a spiky behavior is observed which results from
statistical fluctuations at the energy and magnetization space
boundary. At $T\approx J/k_B$ these fluctuations are still
visible, but on average much smaller than the real minimum at
about $B_{\text{sat}}/3$,\cite{Schroder2005_2} compare
\figref{SuszvsB}. Clearly one can see that the dip in
suszeptibility at about $B_{\text{sat}}/3$ vanishes with
increasing temperature.\cite{Schroder2005_2} The spurious peak
at $T=0$ and $B=0$ reflects the difficulty to obtain the density
of states in close vicinity of the ground state. A comparison
with Metropolis simulations is shown is \figref{SuszvsB}.  The
results obtained using both methods agree well. The fluctuations
in the WL data are due to the missing bins in $g(E,M)$ at the
energy and magnetization space boundaries, since these features
are apparent in all four WL runs.  The statistical sampling
errors are given by the size of the symbols and by the line
widths used in \figref{SuszvsB}.

%===================    figure   =================================
\begin{figure}[ht!]
\centering
\includegraphics[clip,width=0.8\columnwidth]{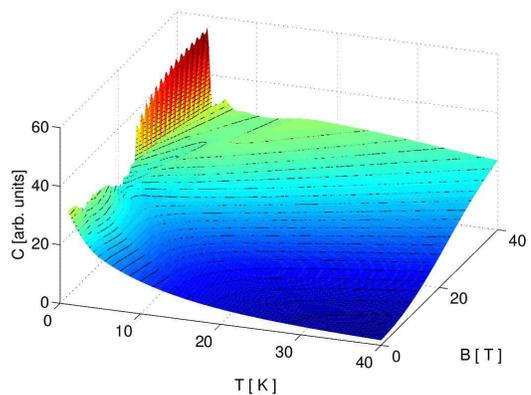}
\caption{\label{SHvsTB} (Color online) Specific heat $C$ as a
function of temperature $T$ and external magnetic field $B$.}
\end{figure}
%===================    figure   =================================

The last function we want to discuss in this paper is the specific
heat $C$ as a function of temperature $T$ and external magnetic
field $B$. As can be seen in \figref{SHvsTB} the specific heat
does not show too strong fluctuations at low temperature and
fields below saturation. It seems that this observable is more
robust against statistical fluctuations of the density of states
than the susceptibility. In order to complete the discussion we
display the specific heat also for magnetic fields above the
saturation field where it shows strong fluctuations. They reflect
magnified inaccuracies at the high energy boundary of $g(E,M)$ and
thus are spurious.

Summarizing, one can say that the proposed self-adaptive version
of the Wang-Landau algorithm can efficiently generate the
density of states $g(E,M)$ of a rather large spin system, such
as the magnetic molecule \{Mo$_{72}$Fe$_{30}$\}. The obvious
advantage is that with one simulation of $g(E,M)$ many
thermodynamic observables can be evaluated as function of
temperature and applied magnetic field. Statistical fluctuations
are apparent in the second derivatives of the two-dimensional
DOS, namely the specific heat and magnetic suszeptibility at low
temperatures. These fluctuations can be minimized at the cost of
computational time, by either increasing the number of bins or
by a more strict flatness criterion. Recently, it was
demonstrated that a fitting of the one-dimensional DOS for a
sampled energy interval by higher order polynomials results in a
less fluctuating specific heat.\cite{Parsons2006} Nevertheless,
it is stated, that this fitting procedure fails if one tries to
extrapolate the DOS outside the sampled energy interval using
the fitting function.  Thus, further improvements are still
needed to provide an effective sampling at the boundaries of
$g(E,M)$. Due to the fact that the density of states varies by
many orders of magnitude it is clear that close to the
boundaries of $g(E,M)$ the statistics becomes poorer.

%%%%%%%%%%%%%%%%%%%%%%%%%%%%%%%%%%%%%%%%%%%%%%%%%%%%%%%%%%%%%%%%%%%%%%%%
\section*{Acknowledgments}

The authors are indebted to D.P.~Landau for having initiated this
work. This work was supported by the Deutsche
Forschungsgemeinschaft (Grant No.  SCHN~615/5-1).

%\bibliography{Wl2D_Fe30_Torbrugge}

\end{document}